\documentclass{osa-article}

\journal{osac}
\articletype{Research Article}
\usepackage{subfigure}
\begin{document}

\title{A Novel Received Signal Strength Assisted Perspective-three-Point Algorithm for Indoor Visible Light Positioning}

\author{Lin Bai,\authormark{1} Yang Yang,\authormark{1} Chunyan Feng,\authormark{1,*}  and Caili Guo\authormark{2}}

\address{\authormark{1}Beijing Key Laboratory of
Network System Architecture and Convergence, School of Information
and Communication Engineering, Beijing University of Posts and Telecommunications\\
\authormark{2}Beijing Laboratory of Advanced Information Networks,
School of Information and Communication Engineering, Beijing University
of Posts and Telecommunications}

\email{\authormark{*}cyfeng@bupt.edu.cn} %% email address is required



\begin{abstract}
In this paper, a received signal strength assisted Perspective-three-Point
positioning algorithm (R-P3P) is proposed for visible light positioning
(VLP) systems. The basic idea of R-P3P is to joint visual and strength
information to estimate the receiver position using 3 LEDs regardless
of the LEDs' orientations. R-P3P first utilizes visual information captured
by the camera to estimate the incidence angles of visible lights.
Then, R-P3P calculates the candidate distances between the LEDs and
the receiver based on the law of cosines and Wu-Ritt's zero decomposition
method. Based on the incidence angles, the candidate distances and
the physical characteristics of the LEDs, R-P3P can select the exact
distances from all the candidate distances. Finally, the linear least
square (LLS) method is employed to estimate the position of the receiver.
Due to the combination of visual and strength information of visible
light signals, R-P3P can achieve high accuracy using 3 LEDs regardless of
the LEDs' orientations. Simulation results show that R-P3P can
achieve positioning accuracy
within 10 cm over 70\% indoor area with low complexity regardless
of LEDs orientations.
\end{abstract}

%%%%%%%%%%%%%%%%%%%%%%%%%%  body  %%%%%%%%%%%%%%%%%%%%%%%%%%
\section{Introduction}
\label{sec:intro} Indoor positioning has attracted
increasing attention recently due to its numerous applications including
indoor navigation, robot movement control and advertisements in shopping
malls. In this research field, visible light positioning (VLP) is
one of the most promising technology due to its high accuracy and
low cost \cite{do2016depth,Pathak2015Visible}. Visible light possesses
strong directionality and low multipath interference, and thus VLP
can achieve high accuracy positioning performance \cite{Pathak2015Visible}.
Besides, VLP utilizes light-emitting diodes (LEDs) as transmitters.
Benefited from the increasing market share of LEDs, VLP has relatively
low cost on infrastructure \cite{Pathak2015Visible}.

VLP typically equips photodiodes (PDs) or cameras as the receiver.
Positioning algorithms using PDs include proximity \cite{gligoric2018visible},
fingerprinting \cite{alam2018accurate} and time of arrival (TOA)
\cite{wang2013TOA}, angle of arrival (AOA) \cite{eroglu2015aoa}
and received signal strength (RSS) \cite{wang2017MDLA,bai2019camera}.
Positioning algorithms using cameras are termed as image sensing \cite{li2018vlc}.
Proximity is the simplest technique, while it only provides proximity
location based on the received signal from a single LED with a unique
identification code. Fingerprinting algorithms can achieve enhanced
accuracy at a high cost for building and updating a database. TOA
and AOA algorithms require complicated hardware
implementation. In contrast,
%RSS algorithms determine the position
%of the receiver based on the power of the received signal from at
%least 3 LEDs. They have the advantage of asynchronous operation and
%simple implementation \cite{do2016depth}. Image sensing algorithms
%for VLP system determine the receiver position by utilizing the geometric
%relations between the LEDs and the camera. They have the advantage
%of simple implementation and high accuracy. From above analysis, we
%can observe that among the algorithms in VLP system,
RSS and image sensing algorithms are the most widely-used methods due to their high
accuracy and moderate cost \cite{do2016depth}. Nowadays, both PD
and the camera are essential parts of smartphones, meaning that RSS
and image sensing algorithms can be easily implemented in such popular
devices \cite{do2016depth}.

However, the RSS and image sensing algorithms also have their own
inherent limitations. In particular, RSS algorithms determine the position of the receiver based on the power of the received signal from at least 3 LEDs, and they have the following
limitations. 1) RSS algorithms limit the orientation of LEDs. Therefore,
in typical RSS algorithms, the LEDs orientations are assumed to face
vertically downwards \cite{bai2019camera,wang2017indoor}. However,
in many scenarios, LEDs do not face vertically
downwards, and thus these RSS algorithms can not be implemented.
%In \cite{wang2017MDLA}, LEDs are assumed to point
%to fixed directions which are known in advance. However,
Besides, it is difficult
to install LEDs according to the default orientation strictly. In
addition, using additional sensors to measure LEDs orientation may induce
measurement errors
\cite{wang2017MDLA}. 2) Besides, RSS algorithms require the orientation
of the receiver to face vertically upward to the ceiling \cite{li2014epsilon},
which is inflexible and slight perturbation of the receiver can affect
the positioning accuracy significantly \cite{bai2019camera}. \cite{li2014epsilon}
exploits additional sensors to measure the receiver orientation. However, it also induces measurement errors, which will further impairs positioning
accuracy.

%On the other hand, based on the normal vector of the receiver
%measured by the sensors, the non-linear least square (NLLS) estimator
%is utilized to obtain the position of the receiver when the receiver
%is not perfectly facing up. This means the starting values of NLLS
%estimator can affect the accuracy \cite{rebaudo2018modelling},
%and the NLLS estimator requires high computation cost.

As for image sensing algorithms, they determine the receiver position
by analyzing the geometric relationship between 3 dimensional (3D)
LEDs and their 2 dimensional (2D) projections on the image plane.
Image sensing algorithms can be classified into two types: single-view
geometry and vision triangulation \cite{do2016depth}. The single-view
geometry methods exploit a single known camera to capture the image
of multiple LEDs \cite{yang2015wearables}, and vision triangulation
methods exploit multiple known cameras to for 3D position measurement
\cite{rahman2011high}. Nowadays, mobile devices with one front camera
occupy a large market share. Therefore, single-view geometry methods
are more suitable for indoor positioning. Perspective-n-point (PnP)
is a typical single-view geometry algorithm that has been extensively
studied \cite{li2018vlc,lepetit2009epnp,kneip2011novel}. However,
PnP algorithms require at least 4 LEDs to obtain a deterministic 3D
position \cite{lepetit2009epnp}.
%Therefore, the application of PnP
%algorithms is limited in scenarios where deployed with insufficient
%LEDs or the camera hard to capture enough LEDs.

To address the problems in both the RSS and the PnP algorithms, in
our previous work \cite{bai2019camera}, we proposed a camera-assisted
received signal strength ratio algorithm (CA-RSSR). CA-RSSR exploits
both the strength and visual information of visible lights and it
achieves centimeter-level 2D positioning accuracy with 3 LEDs regardless
of the receiver orientation without any additional sensors. However,
CA-RSSR still requires LEDs to face vertically downwards. Besides,
CA-RSSR uses the NLLS method for positioning, which means the
accuracy depends on the starting values of the NLLS estimator and
the NLLS method increases the complexity. In addition, CA-RSSR
requires at least 5 LEDs to achieve 3D positioning, which
is even worse than PnP algorithms. Therefore, the VLP algorithm which
can be widely used still remains to be developed.

Against the aforementioned background, we propose a novel RSS assisted
Perspective-three-Point algorithm (R-P3P) that can be widely used for indoor scenarios. First, R-P3P exploits the visual information captured
by the camera to estimate the incidence angles of the visible light
based on the single-view geometry. Then, R-P3P estimate the candidate
distances between the LEDs and the receiver based on the law of cosines
and Wu-Ritt's zero decomposition method. Based on the candidate
distances, the estimated incidence angles and the semi-angles of the
LEDs, the irradiance angles of the visible light can be obtained by
the strength information captured by the PD, and then the distances
between the LEDs and the receiver can be determined. Finally, based
on the distances, the position of
the receiver can be obtained by the linear least square (LLS) method.
Therefore, compared with CA-RSSR, R-P3P can mitigate the limitation
of LEDs orientation.
%due to the combination of the strength and visual
%information of visible lights.
Besides, the LLS method can
avoid the potential side effect of the starting values of the NLLS
method and requires lower computation cost than the NLLS method.
%Meanwhile, the
%computation complexity of R-P3P is low due to the use of the LLS method.
On the other hand, compared with the PnP algorithms, R-P3P only requires
3 LEDs for 3D positioning. Therefore, the algorithm can be more widely-used
for indoor positioning. Simulation results show that R-P3P can achieve
positioning accuracy
within 10 cm over 70\% indoor area with low complexity regardless of
LEDs orientations.

The rest of the paper is organized as follows. Section \ref{sec:SM}
introduces the system model. The proposed positioning algorithm is
detailed in Section \ref{sec:CT-CA-RSSR}. Simulation results are
presented in Section \ref{sec:simulation}. Finally, the paper is
concluded in Section \ref{sec:CONCLUSION}.\vspace{-0cm}

\section{System Model}

\label{sec:SM}
\begin{figure}
\begin{centering}
\includegraphics[scale=0.55]{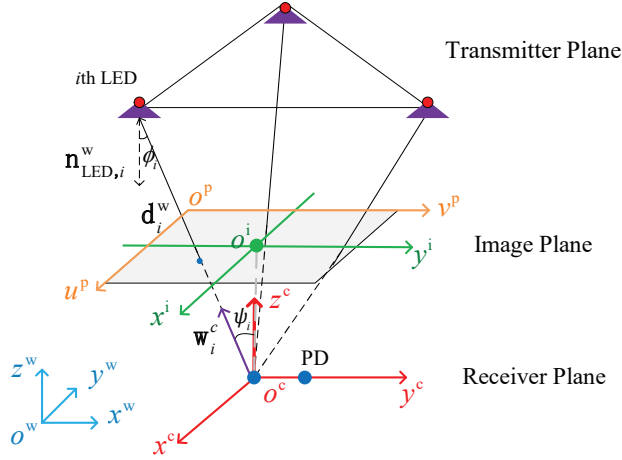}
\par\end{centering}
\caption{\label{fig:A-system-block}The system diagram of the VLP system.}
\end{figure}

The system diagram is illustrated in Fig. \ref{fig:A-system-block}.
Four coordinate systems are utilized for positioning, which are the
pixel coordinate system (PCS) $o^{\textrm{p}}-u^{\textrm{p}}v^{\textrm{p}}$
on the image plane, the image coordinate system (ICS) $o^{\textrm{i}}-x^{\textrm{i}}y^{\textrm{i}}$
on the image plane, the camera coordinate system (CCS) $o^{\textrm{c}}-x^{\textrm{c}}y^{\textrm{c}}z^{\textrm{c}}$
and the world coordinate system (WCS) $o^{\textrm{w}}-x^{\textrm{w}}y^{\textrm{w}}z^{\textrm{w}}$.
As shown in Fig. \ref{fig:A-system-block}, different colors represent different coordinate systems.
In PCS, ICS and CCS, the axes $u^{\textrm{p}}$, $x^{\textrm{i}}$
and $x^{\textrm{c}}$ are parallel to each other and, similarly, $v^{\textrm{p}}$,
$y^{\textrm{i}}$ and $y^{\textrm{c}}$ are also parallel to each
other.
Besides, $o^{\textrm{p}}$ is in the upper left corner of the image plane and $o^{\textrm{i}}$ is in the center of the image plane.
In addition, $o^{\textrm{i}}$ is termed as the principal point, whose pixel coordinate is $\left(u_{0},v_{0}\right)$. In contrast, $o^{\textrm{c}}$ is termed as the camera optical center.
Furthermore, $o^{\textrm{i}}$ and $o^{\textrm{c}}$ are on the optical axis.
The distance between $o^{\textrm{c}}$ and $o^{\textrm{i}}$
is the focal length $f$, and thus the $z$-coordinate of the image
plane in CCS is $z^{\mathrm{c}}=f$.

In the proposed positioning system, %$K\left(K=3\right)$
3 LEDs are the transmitters mounted on the ceiling. The receiver is composed of a PD and a standard pinhole camera, and they are close to each other. As shown in Fig.
\ref{fig:A-system-block}, $\mathbf{n}_{\textrm{LED},i}^{\textrm{w}}$
denotes the unknown unit normal vector of the $i$th LED in the WCS. Besides, $\mathbf{s}_{i}^{\textrm{w}}=\left(x_{i}^{\textrm{w}},y_{i}^{\textrm{w}},z_{i}^{\textrm{w}}\right)$
($i\in\left\{ 1,2,3\right\} $) is the coordinate of the $i$th
LED in the WCS, which are assumed to be known at the transmitter and
can be obtained by the receiver through visible light communications
(VLC). In contrast, $\mathbf{r}^{\textrm{w}}=\left(x_{r}^{\textrm{w}},y_{r}^{\textrm{w}},z_{r}^{\textrm{w}}\right)$
is the world coordinate of the receiver to be positioned. In addition,
$\phi_{i}$ and $\psi_{i}$ are the irradiance angle and the incidence
angle of the visible lights, respectively. Furthermore, $\mathbf{w}_{i}^{\textrm{c}}$
and $\mathbf{d}_{i}^{\textrm{w}}$ denote the vectors from the receiver
to the $i$th LED in the CCS and the WCS, respectively.
%In the pinhole camera, the $i$th LED, the projection of the $i$th LED onto the image plane
%and the camera optical center $o^{\textrm{c}}$ are on the same straight line.
%The original point of ICS, $o^{\textrm{i}}$, is termed as the principal
%point, whose pixel coordinate is $\left(u_{0},v_{0}\right)$. The pixel coordinate of the projection of the $i$th LED is denoted by $\mathbf{s}_{i}^{\textrm{p}}=\left(u_{i}^{\textrm{p}},v_{i}^{\textrm{p}}\right)$.

LEDs with Lambertian radiation pattern are considered. The line of
sight (LoS) link is the dominant component in the optical channel,
and thus this work only considers the LoS channel for simplicity \cite{Komine2004Fundamental}.
The channel direct current (DC) gain between the $i$th LED and the
PD is given by \cite{yang2019relay}
\begin{equation}
H_{i}=\frac{\left(m+1\right)A}{2\pi d_{i}^{2}}\cos^{m}\left(\phi_{i}\right)T_{s}\left(\psi_{i}\right)g\left(\psi_{i}\right)\cos\left(\psi_{i}\right)\label{eq:1}
\end{equation}
where $m$ is the Lambertian order of the LED, given by $m=\frac{-\ln2}{\ln\left(\cos\Phi_{\frac{1}{2}}\right)}$, where
$\Phi_{\frac{1}{2}}$ denotes the semi-angles of the LED. In addition,
$d_{i}=\left\Vert \mathbf{d}_{i}^{\textrm{w}}\right\Vert =\left\Vert \mathbf{s}_{i}^{\textrm{w}}-\mathbf{r}^{\textrm{w}}\right\Vert $,
where $\left\Vert \cdot\right\Vert $ denotes Euclidean norm of vectors,
$A$ is the physical area of the detector at the PD, $T_{s}\left(\psi_{i}\right)$
is the gain of the optical filter, and $g\left(\psi_{i}\right)=\begin{cases}
\frac{n^{2}}{\sin^{2}\Psi_{c}}, & 0\leq\psi_{i}\leq\Psi_{c}\\
0, & \psi_{i}\geq\Psi_{c}
\end{cases}$ is the gain of the optical concentrator, where $n$ is the refractive
index of the optical concentrator and $\Psi_{c}$ is the field of
view (FoV) of the PD. The received optical power from the $i$th LED
can be expressed as
\begin{equation}
P_{r,i}=P_{t}H_{i}=\frac{C}{d_{i}^{2}}\cos^{m}\left(\phi_{i}\right)\cos\left(\psi_{i}\right)\label{eq:3}
\end{equation}
where $P_{t}$ denotes the optical power of the LEDs and $C=P_{t}\frac{\left(m+1\right)A}{2\pi}T_{s}\left(\psi_{i}\right)g\left(\psi_{i}\right)$
is a constant. The signal-to-noise ratio (SNR) is calculated as $SNR_{i}=10\log_{10}\frac{\left(P_{r,i}R_{p}\right)^{2}}{\sigma_{\mathrm{noise},i}^{2}}$,
where $R_{p}$ is the efficiency of the optical to electrical conversion
and $\sigma_{\mathrm{noise},i}^{2}$ means the total noise variance.

\vspace{-0cm}

\section{\label{sec:CT-CA-RSSR}Received Signal Strength Assisted Perspective-three-Point Algorithm (R-P3P)}
In this section, a novel visible light positioning algorithm, termed
as R-P3P is proposed. R-P3P mainly consists of three steps. In the first step,
the incidence angle is estimated according to the visual information
captured by the camera based on the single-view geometry. Then, the
candidate distances between the LEDs and the receiver is obtained
based on the law of cosines and Wu-Ritt\textquoteright s zero
decomposition method \cite{gao2003complete}. Next, based on the candidate
distances, the incidence angles and the semi-angles of the LEDs, the
irradiance angles are calculated utilizing the RSS received by the
PD and then the exact distances between the LEDs and the receiver
can be obtained. Finally, based on the distances, the position of
the receiver is estimated by the LLS algorithm.\vspace{-0cm}

\subsection{\label{subsec:Incidence-Angle-Estimation}Incidence Angle Estimation}

In the pinhole camera, the pixel coordinate of the projection of the $i$th LED is denoted by $\mathbf{s}_{i}^{\textrm{p}}=\left(u_{i}^{\textrm{p}},v_{i}^{\textrm{p}}\right)$,
and this coordinate can be obtained by the camera through image processing
\cite{li2018vlc}.
Based on the single-view geometry theory, the $i$th LED, the projection of the $i$th LED onto the image plane and $o^{\textrm{c}}$
are on the same straight line. Therefore, the camera coordinates of the $i$th LED can be expressed as follows
\begin{equation}
\mathbf{s}_{i}^{\textrm{c}}=\begin{bmatrix}x_{i}^{\textrm{c}}\\
y_{i}^{\textrm{c}}\\
z_{i}^{\textrm{c}}
\end{bmatrix}=\mathbf{M^{-1}}\cdot z_{i}^{\textrm{c}}\begin{bmatrix}u_{i}^{\textrm{p}}\\
v_{i}^{\textrm{p}}\\
1
\end{bmatrix}\label{eq:18}
\end{equation}
where $\mathbf{M}=\begin{bmatrix}f_{u} & 0 & u_{0}\\
0 & f_{v} & v_{0}\\
0 & 0 & 1
\end{bmatrix}$ is the intrinsic parameter matrix of the camera, which can be calibrated
in advance \cite{kneip2011novel}. Besides, $f_{u}=\frac{f}{d_{x}}$
and $f_{v}=\frac{f}{d_{y}}$ denote the focal ratio along
$u$ and $v$ axes in pixels, respectively. In addition, $d_{x}$
and $d_{y}$ are the physical size of each pixel in the $x$ and $y$
directions on the image plane, respectively.

In CCS, the vector from $o^{\textrm{c}}$ to the $i$th LED, $\mathbf{w}_{i}^{\textrm{c}}$,
can be expressed as
\begin{equation}
\mathbf{w}_{i}^{\textrm{c}}=\mathbf{s}_{i}^{\textrm{c}}-\mathbf{o}^{\textrm{c}}=\left(x_{i}^{\textrm{c}},y_{i}^{\textrm{c}},z_{i}^{\textrm{c}}\right)\label{eq:19}
\end{equation}
where $\mathbf{o}^{\textrm{c}}=\left(0^{\textrm{c}},0^{\textrm{c}},0^{\textrm{c}}\right)$
is the origin of the camera coordinate. The estimated incidence angle
of the $i$th LED can be calculated as
\begin{equation}
\psi_{i,\textrm{est}}=\arccos\frac{\mathbf{w}_{i}^{\textrm{c}}\cdot\left(\mathbf{n}_{\textrm{cam}}^{\textrm{c}}\right)^{\textrm{T}}}{\left\Vert \mathbf{w}_{i}^{\textrm{c}}\right\Vert }\label{eq:8}
\end{equation}
where $\mathbf{n}_{\textrm{cam}}^{\textrm{c}}=\left(0^{\textrm{c}},0^{\textrm{c}},1^{\textrm{c}}\right)$
is the unit normal vector of the camera in CCS and is known at
the receiver side. Besides, $\left(\cdot\right)^{\textrm{T}}$ denotes
the transposition of matrices. Since the absolute value of $\psi_{i,\textrm{est}}$
remains the same in different coordinate systems, the estimated incidence
angles in WCS are also given by (\ref{eq:8}). In this way, R-P3P
is able to obtain the incidence angles regardless of the receiver
orientation.\vspace{-0cm}

\subsection{\label{subsec:Dis-Estimation}Distance Estimation}

\begin{figure}
\begin{centering}
\includegraphics[scale=0.8]{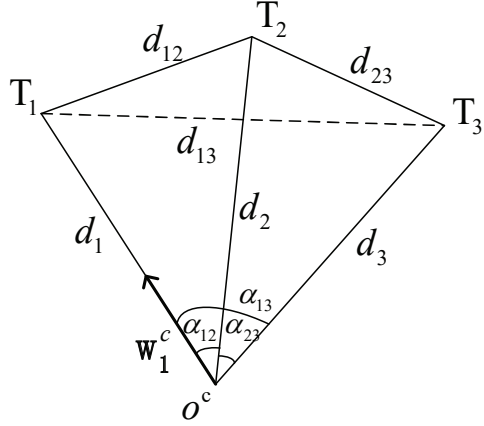}
\par\end{centering}
\caption{\label{fig:cosine theorem}The geometrical relationship among LEDs and the camera optical center for the utilization of
the law of cosines.}
\end{figure}

Figure \ref{fig:cosine theorem} shows the geometric relations among LEDs and the camera. As shown in Fig. \ref{fig:cosine theorem},
$\textrm{T}_{i}$ ($i\in\left\{ 1,2,3\right\} $) is the $i$th
LED and $o^{\textrm{c}}$ is the camera optical center. The distance
between $\textrm{T}_{i}$ and $\textrm{T}_{j}$, $d_{ij}$ ($i,j\in\left\{ 1,2,3\right\} ,i\neq j$),
is known in advance. Besides, $\mathbf{w}_{i}^{\textrm{c}}$ ($i\in\left\{ 1,2,3\right\} $),
which can be calculated by (\ref{eq:19}), are the vectors from the
receiver to $\textrm{T}_{i}$ in CCS. Furthermore, $\alpha_{ij}$
($i,j\in\left\{ 1,2,3\right\} ,i\neq j$) is the angle between
$\mathbf{w}_{i}^{\textrm{w}}$ and $\mathbf{w}_{j}^{\textrm{w}}$,
i.e., $\alpha_{ij}=\angle\textrm{T}_{i}o^{\textrm{c}}\textrm{T}_{j}$,
which can be calculated as
\begin{equation}
\alpha_{ij}=\arccos\frac{\mathbf{w}_{i}^{\textrm{c}}\cdot\left(\mathbf{w}_{j}^{\textrm{c}}\right)^{\textrm{T}}}{\left\Vert \mathbf{w}_{i}^{\textrm{c}}\right\Vert \left\Vert \mathbf{w}_{j}^{\textrm{c}}\right\Vert }.\label{eq:25}
\end{equation}
We define $\triangle\textrm{T}_{i}o^{\textrm{c}}\textrm{T}_{j}$ as
the triangle constructed by the vertices $\textrm{T}_{i}$, $o^{\textrm{c}}$
and $\textrm{T}_{j}$. According to the law of cosines, in the triangle
$\triangle\textrm{T}_{i}o^{\textrm{c}}\textrm{T}_{j}$, we have
\begin{equation}
d_{i}^{2}+d_{j}^{2}-2d_{i}d_{j}\cos\alpha_{ij}=d_{ij}^{2}.\label{eq:26}
\end{equation}
To simplify (\ref{eq:26}), let
\begin{equation}
\begin{cases}
r=2\cos\alpha_{12}\\
q=2\cos\alpha_{13}\\
q=2\cos\alpha_{23},
\end{cases}\label{eq:27}
\end{equation}
\begin{equation}
\begin{cases}
d_{1}=xd_{3}\\
d_{2}=yd_{3},
\end{cases}\label{eq:28}
\end{equation}
and
\begin{equation}
\begin{cases}
d_{12}^{2}=vd_{3}^{2}\\
d_{23}^{2}=ad_{12}^{2}=avd_{3}^{2}\\
d_{13}^{2}=bd_{12}^{2}=bvd_{3}^{2}.
\end{cases}\label{eq:29}
\end{equation}
Since $d_{3}\neq0$, we can obtain the following equation system which
is equivalent to (\ref{eq:26})
\begin{equation}
\begin{cases}
v=x^{2}+y^{2}-xyr\\
bv=x^{2}+1-xq\\
av=1+y^{2}-yp.
\end{cases}\label{eq:31}
\end{equation}
Since $r<2$, we have $v=x^{2}+y^{2}-xyr>0$. Thus, $d_{3}$ can be
uniquely determined by $d_{3}=\frac{d_{12}}{\sqrt{v}}$, where
$v$ requires to be calculated. Besides, we can eliminate $v$ from
(\ref{eq:31}), and thus we have
\begin{equation}
\begin{cases}
\left(1-a\right)y^{2}-ax^{2}+axyr-yp+1=0\\
\left(1-b\right)x^{2}-by^{2}+bxyr-xq+1=0.
\end{cases}\label{eq:32}
\end{equation}
Following the same method in \cite{gao2003complete}, $d_{i}$ ($i\in\left\{ 1,2,\ldots,K\right\} $)
can be obtianed by solving (\ref{eq:32}) based on Wu-Ritt's zero
decomposition method \cite{Wu1984basic} as follows
\begin{equation}
\begin{cases}
d_{3}=\frac{d_{12}}{\sqrt{v}}\\
d_{1}=xd_{3}\\
d_{2}=yd_{3}.
\end{cases}\label{eq:34}
\end{equation}
As the same with \cite{gao2003complete}, there four groups of $d_{i}$
($i\in\left\{ 1,2,\ldots,K\right\} $). The typical P3P methods require
the fourth beacon to obtain the right solution of $d_{i}$ \cite{gao2003complete,kneip2011novel,lepetit2009epnp}.
In contrast, we obtain the right solution based on the RSS captured
by the PD in the next subsection.

\subsection{\label{subsec:irradiance angle}Irradiance Angle Estimation}

According to (\ref{eq:3}), the RSS captured by the PD from the $i$th
LED can be expressed as
\begin{equation}
P_{r,i}=\frac{C}{d_{i}^{2}}\cos^{m}\left(\phi_{i}\right)\cos\left(\psi_{i}\right).\label{eq:5}
\end{equation}
Since the distance between the PD and the camera, $d_{\textrm{PC}}$,
is much smaller than the distances between the LEDs and the receiver,
we omit $d_{\textrm{PC}}$ in the algorithm. However, the effect of
$d_{\textrm{PC}}$ on R-P3P's performance will be evaluated in the simulations.
Therefore, with the incidence angle estimated by (\ref{eq:8}), we
can obtain the irradiance angle $\phi_{i}$ ($i\in\left\{ 1,2,3\right\} $)
as follows
\begin{equation}
\cos\left(\phi_{i}\right)=\left(\frac{P_{r,i}\cdot d_{i}^{2}}{C\cdot\cos\left(\psi_{i,\mathrm{est}}\right)}\right)^{\frac{1}{m}}.\label{eq:33}
\end{equation}
With the four groups of $d_{i}$ ($i\in\left\{ 1,2,3\right\} $) obtained by (\ref{eq:34}),
we can obtain four groups of $\phi_{i}$ ($i\in\left\{ 1,2,3\right\} $).
Fortunately, the semi-angles of the LEDs, $\Phi_{\frac{1}{2}}$,
are known in advance. This means that the right solution of $\phi_{i}$
($i\in\left\{ 1,2,3\right\} $) have to comply with the following
constraints
\begin{equation}
\cos\left(\Phi_{\frac{1}{2}}\right)\leq\cos\left(\phi_{i}\right)\leq\cos\left(\frac{\pi}{2}\right).\label{eq:35}
\end{equation}
We can estimate $\phi_{i}$ ($i\in\left\{ 1,2,3\right\} $)
by (\ref{eq:35}).
%However, there may be more than one groups of $\phi_{i}$
%($i\in\left\{ 1,2,3\right\} $) comply (\ref{eq:35}).
However, consider the effect of noise and $d_{\textrm{PC}}$,
there may be no group of $\phi_{i}$
($i\in\left\{ 1,2,3\right\} $) comply (\ref{eq:35}) or there may be more than one groups of $\phi_{i}$
($i\in\left\{ 1,2,3\right\} $) comply (\ref{eq:35}).
For the former case,
%that there is no appropriate $\phi_{i}$ ($i\in\left\{ 1,2,3\right\} $),
we give a tolerance for (\ref{eq:35}) with the step of 5\% until
we find out one group of exact $\phi_{i}$ ($i\in\left\{ 1,2,3\right\} $).
For the latter case,
%that there are more than one groups of $\phi_{i}$ ($i\in\left\{ 1,2,3\right\} $),
we choose the final $\phi_{i}$ ($i\in\left\{ 1,2,3\right\} $)
from all the groups that comply (\ref{eq:35}) randomly. These measures will undoubtedly introduce
positioning errors. Fortunately, the probability of these cases
is very low, and thus the accuracy of R-P3P is almost the same with
the typical PnP method that requires 4 LEDs, which will be shown in the simulations.

Based on the estimated $\phi_{i}$ ($i\in\left\{ 1,2,3\right\} $),
$d_{i}$ ($i\in\left\{ 1,2,3\right\} $) can be further determined.
In this way, we can estimate the distances between the LEDs and the
receiver using only three LEDs.

\vspace{-0cm}

\subsection{\label{subsec:2D-CT}Position Estimation By Linear Least Square Algorithm}

The distances between the LEDs and the receiver obtained in \ref{subsec:irradiance angle} can be expressed as follows
\begin{equation}
\begin{cases}
d_{1}=\left\Vert \mathbf{s}_{1}^{\textrm{w}}-\mathbf{r}^{\textrm{w}}\right\Vert _{\textrm{est}}\\
d_{2}=\left\Vert \mathbf{s}_{2}^{\textrm{w}}-\mathbf{r}^{\textrm{w}}\right\Vert _{\textrm{est}}\\
d_{3}=\left\Vert \mathbf{s}_{3}^{\textrm{w}}-\mathbf{r}^{\textrm{w}}\right\Vert _{\textrm{est}}\text{.}
\end{cases}\label{eq:61}
\end{equation}
In practice, LEDs are usually deployed at the same height (i.e., $z_{1}^{\textrm{w}}=z_{2}^{\textrm{w}}=z_{3}^{\textrm{w}}$) and hence
R-P3P can estimate the 2D position of the receiver $\left(x_{r}^{\textrm{w}},y_{r}^{\textrm{w}}\right)$ based on the following standard LLS estimator
\begin{equation}
\mathbf{\hat{X}=(A^{\mathrm{T}}A)^{\mathrm{-1}}A^{\mathrm{T}}b}.\label{eq:53}
\end{equation}
where $\hat{\mathbf{X}}=\begin{bmatrix}x_{r,\textrm{est}}^{\textrm{w}}\\
y_{r,\textrm{est}}^{\textrm{w}}\end{bmatrix}$ is the estimate of $\mathbf{X}=\begin{bmatrix}x_{r}^{\textrm{w}}\\
y_{r}^{\textrm{w}}
\end{bmatrix}$. Besides,
%using two linear equations. These linear equations can be simply obtained
%by subtracting the second and the third equations from the first one
%in (\ref{eq:61}), which can be expressed in a matrix form as follows:
%\begin{equation}
%\mathbf{AX=b}\label{eq:47}
%\end{equation}
%where
%\begin{equation}
%\mathbf{X}=\begin{bmatrix}x_{r}^{\textrm{w}}\\
%y_{r}^{\textrm{w}}
%\end{bmatrix},\label{eq:50}
%\end{equation}
\begin{equation}
\mathbf{A}=\begin{bmatrix}x_{2}^{\textrm{w}}-x_{1}^{\textrm{w}} & y_{2}^{\textrm{w}}-y_{1}^{\textrm{w}}\\
x_{3}^{\textrm{w}}-x_{1}^{\textrm{w}} & y_{3}^{\textrm{w}}-y_{1}^{\textrm{w}}
\end{bmatrix},\label{eq:48}
\end{equation}
and
\begin{equation}
\mathbf{b}=\frac{1}{2}\begin{bmatrix}C_{1}^{2}-C_{2}^{2}+\left(x_{2}^{\textrm{w}}\right)^{2}+\left(y_{2}^{\textrm{w}}\right)^{2}-\left(x_{1}^{\textrm{w}}\right)^{2}-\left(y_{1}^{\textrm{w}}\right)^{2}\\
C_{1}^{2}-C_{3}^{2}+\left(x_{3}^{\textrm{w}}\right)^{2}+\left(y_{3}^{\textrm{w}}\right)^{2}-\left(x_{1}^{\textrm{w}}\right)^{2}-\left(y_{1}^{\textrm{w}}\right)^{2}
\end{bmatrix}.\label{eq:52}
\end{equation}
%Obviously, the equations apply to a standard LLS estimator given by
%\begin{equation}
%\mathbf{\hat{X}=(A^{\mathrm{T}}A)^{\mathrm{-1}}A^{\mathrm{T}}b}.\label{eq:53}
%\end{equation}
%where $\hat{\mathbf{X}}$ is the estimate of $\mathbf{X}$.

Since $z_{1}^{\textrm{w}}=z_{2}^{\textrm{w}}=z_{3}^{\textrm{w}}$, $z$-coordinate of the receiver can be calculated by substituting (\ref{eq:53})
into the first equation of (\ref{eq:61}), which can be expressed
as follows
\begin{equation}
z_{r,\textrm{est}}^{\textrm{w}}=z_{1}^{\textrm{w}}\pm\Delta\label{eq:66}
\end{equation}
where $\Delta=\sqrt{C_{1}^{2}-\left(x_{1}^{\textrm{w}}-x_{r,\textrm{est}}^{\textrm{w}}\right)^{2}-\left(y_{1}^{\textrm{w}}-y_{r,\textrm{est}}^{\textrm{w}}\right)^{2}}$.
Since $H_{i}$ is the quadratic of $d_{i}$, as shown in (\ref{eq:1}),
we can obtain two $z$-coordinates of the receiver. However, the ambiguous
solution, $z_{r,\textrm{est}}^{\textrm{w}}=h+\Delta$, can be easily
eliminated as it implies the height of the receiver is beyond the
ceiling. Therefore, R-P3P can determine the 3D position of the receiver,
$\mathbf{r}_{\mathrm{est}}^{\mathrm{w}}=\left(x_{r,\textrm{est}}^{\textrm{w}},y_{r,\textrm{est}}^{\textrm{w}},z_{r,\textrm{est}}^{\textrm{w}}\right)$,
by only 3 LEDs with the LLS method.

\global\long\def\arraystretch{0.9}%
\begin{table}[t]
\centering{}\centering{}\caption{\label{tab:Parameters-used-for}System Parameters.}
\begin{tabular}{>{\raggedright}m{5.5cm}|>{\centering}m{2.5cm}}
\hline
{\footnotesize{}{}{}{}{}{}Parameter}  & {\footnotesize{}{}{}{}{}{}Value}\tabularnewline
\hline
{\footnotesize{}{}{}{}{}{}Room size ($\textrm{length}\times\textrm{width}\times\textrm{height}$)}  & {\footnotesize{}{}{}{}{}{}$5\,\mathrm{m}\times5\,\mathrm{m}\times3\,\mathrm{m}$}\tabularnewline
\hline
{\footnotesize{}{}{}{}{}{}LED coordinates}  & {\footnotesize{}{}{}{}{}$\left(2,2,3\right)$, $\left(2,3,3\right)$,}{\footnotesize\par}

{\footnotesize{}{}{}{}{}$\left(3,3,3\right)$, $\left(3,2,3\right)$
$(2.5,2.5,3)$}\tabularnewline
\hline
{\footnotesize{}{}{}{}{}{}LED transmit optical power, $P_{t}$}  & {\footnotesize{}{}{}{}{}{}2.2 $\textrm{W}$}\tabularnewline
\hline
{\footnotesize{}{}{}{}{}{}LED semi-angle, $\Phi_{\frac{1}{2}}$}  & {\footnotesize{}{}{}{}{}{}$60{^{\circ}}$}\tabularnewline
\hline
{\footnotesize{}{}{}{}{}{}PD detector physical area, $A$}  & {\footnotesize{}{}{}{}{}{}1 ${\textstyle \mathrm{cm^{2}}}$}\tabularnewline
\hline
{\footnotesize{}{}{}{}{}{}Gain of the optical filter, $T_{s}$}  & {\footnotesize{}{}{}{}{}{}1}\tabularnewline
\hline
{\footnotesize{}{}{}{}{}{}Refractive index of the optical concentrator,
$n$}  & {\footnotesize{}{}{}{}{}{}1.5}\tabularnewline
\hline
{\footnotesize{}{}{}{}{}{}Receiver FoV, $\Psi_{c}$}  & {\footnotesize{}{}{}{}{}{}$60{^{\circ}}$}\tabularnewline
\hline
{\footnotesize{}{}{}Distance between the PD and the camera, $d_{\textrm{pc}}$}  & {\footnotesize{}{}{}$1\,\textrm{cm}$}\tabularnewline
\hline
\end{tabular}
\end{table}

\vspace{-0cm}

\section{\label{sec:simulation}SIMULATION RESULTS AND ANALYSES}

As R-P3P simultaneously utilizes visual and strength information,
a typical PnP algorithm \cite{gao2003complete} and CA-RSSR \cite{bai2019camera}
are conducted as the baseline schemes in this section. The PnP algorithm
utilizes the visual information only. Besides, CA-RSSR exploits both
visual and strength information.

The system parameters are listed in Table \ref{tab:Parameters-used-for}.
Assume that visible light signals are modulated by on-off keying (OOK).
All statistical results are averaged over $10^{5}$ independent runs.
For each simulation run, the receiver positions are selected in the
room randomly. To reduce the error caused by the channel noise, the
received optical power is calculated as the average of 1000 measurements
\cite{wang2017MDLA}. The pinhole camera is calibrated and has a principal
point $\left(u_{0},v_{0}\right)=\left(320,240\right)$, and a
focal ratio $f_{u}=f_{v}=800$. The image noise is modeled as a white
Gaussian noise having an expectation of zero and a standard deviation
of $2$ pixels \cite{zhou2019robust}. Since
the image noise affects the pixel coordinate of the LEDs'
projection on the image plane, the pixel coordinate is obtained
by processing 10 images for the same position.

We evaluate the performance of R-P3P in terms of its coverage, accuracy
and computational cost in the 3D-positioning case. We define coverage
ratio (CR) of the positioning algorithms as
\begin{equation}
CR=\frac{A_{\textrm{effective}}}{A_{\textrm{total}}}\label{eq:25-1}
\end{equation}
where $A_{\textrm{effective}}$ is the indoor area where the algorithm
is feasible and $A_{\textrm{total}}$ is the entire indoor area. Besides,
the positioning error (PE) is used to quantify the accuracy performance
which is defined as
\begin{equation}
PE=\left\Vert \mathbf{r}_{\textrm{true}}^{\textrm{w}}-\mathbf{r}_{\textrm{est}}^{\textrm{w}}\right\Vert \label{eq:25-2}
\end{equation}
where $\mathbf{r}_{\textrm{true}}^{\textrm{w}}=\left(x_{r,\textrm{true}}^{\textrm{w}},y_{r,\textrm{true}}^{\textrm{w}},z_{r,\textrm{true}}^{\textrm{w}}\right)$
and $\mathbf{r_{\textrm{est}}^{\textrm{w}}}=\left(x_{r,\textrm{est}}^{\textrm{w}},y_{r,\textrm{est}}^{\textrm{w}},z_{r,\textrm{est}}^{\textrm{w}}\right)$
are the world coordinates of the actual and estimated positions of
the receiver, respectively. Furthermore, we exploit the execution
time to evaluate the computational cost.

\global\long\def\arraystretch{0.9}%
\begin{table}
\centering{}\centering{}\caption{\label{tab:CR}The Required Number of LEDs for The Positioning Schemes.}

\centering{}%
\begin{tabular}{>{\centering}m{2.5cm}|>{\centering}m{3.5cm}}
\hline
{\footnotesize{}{}{}Positioning Scheme}  & {\footnotesize{}{}{}Sufficient Number of LEDs}\tabularnewline
\hline
{\footnotesize{}{}{}PnP}  & {\footnotesize{}{}{}4}\tabularnewline
\hline
{\footnotesize{}{}{}CA-RSSR}  & {\footnotesize{}{}{}5}\tabularnewline
\hline
{\footnotesize{}{}{}R-P3P}  & {\footnotesize{}{}{}3}\tabularnewline
\hline
\end{tabular}
\end{table}

\vspace{-0cm}

\subsection{Coverage Performance Of R-P3P}

\begin{figure}[htbp]
\centering
\subfigure[LEDs tilt with 0\textdegree.]{\label{Fig:R1}\includegraphics[scale=0.65]{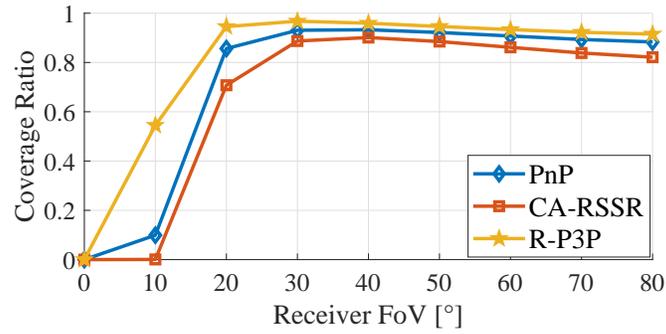}} \quad{}\subfigure[LEDs tilt with 10\textdegree.]{\label{Fig:R2} \includegraphics[scale=0.65]{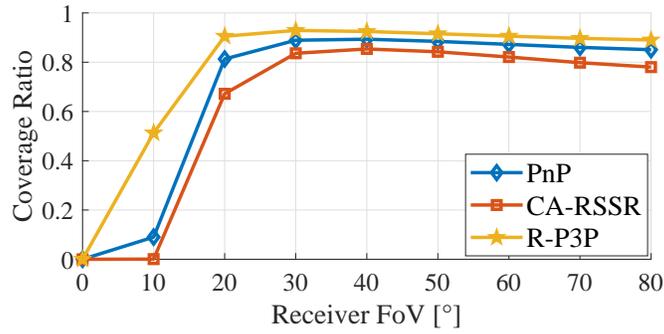}} \quad{}\subfigure[LEDs tilt with 30 \textdegree.]{\label{Fig:R3}\includegraphics[scale=0.65]{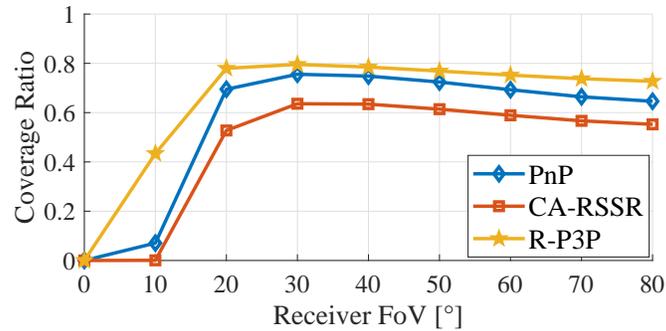}}\\

\caption{The comparison of the 3D-positioning CR performance among R-P3P, CA-RSSR
and the PnP algorithm with varying FoVs of the receiver.}
%\subref{Fig:R1}}
 \label{fig:CR-random-far}
\end{figure}

Table \ref{tab:CR} provides the required number of LEDs for 3D positioning
for R-P3P, CA-RSSR and the PnP algorithm. As we can observe, R-P3P
requires the least number of LEDs. Figure \ref{fig:CR-random-far}
shows the comparisons of the coverage ratio (CR) performance among
the three algorithms with the FoVs, $\Psi_{c}$, varying from $0{^{\circ}}$
to $80{^{\circ}}$. Besides, the LEDs tilt with a angle $\theta=0{^{\circ}}$,
$\theta=10{^{\circ}}$ and $\theta=30{^{\circ}}$ for Fig. \ref{Fig:R1},
Fig. \ref{Fig:R2} and Fig. \ref{Fig:R3}, respectively. The positioning
samples are chosen along the length, width and height of the room,
with a five centimeters separation from each other. A SNR of 13.6
dB is assumed according to the reliable communication requirement
of OOK modulation \cite{Komine2004Fundamental}. As shown in Fig.
\ref{fig:CR-random-far}, R-P3P achieves the highest CR for all $\Psi_{c}$
regardless of $\theta$. It performs consistently well from $\Psi_{c}=20{^{\circ}}$
to $\Psi_{c}=80{^{\circ}}$ with the CR exceeding 90\% for $\theta=0{^{\circ}}$
and $\theta=10{^{\circ}}$, and the CR exceeding 70\% for $\theta=30{^{\circ}}$.
The CR of R-P3P is more than 2\% , 3\% and 5\% higher than the PnP
algorithm for $\theta=0{^{\circ}}$, $\theta=10{^{\circ}}$ and $\theta=30{^{\circ}}$,
respectively. Meanwhile, the CR of R-P3P is more than 8\% , 10\% and
18\% higher than the CA-RSSR for $\theta=0{^{\circ}}$, $\theta=10{^{\circ}}$
and $\theta=30{^{\circ}}$, respectively. As we can observe from Fig. \ref{fig:CR-random-far},
as the tilt angle of the LEDs increases, the CR for all the three
algorithms decreases, and the CR performance advantage of R-P3P compared
with the other two algorithms increases. Besides, the CR of R-P3P
is more than 40\% for all the three $\theta$ for $\Psi_{c}=10{^{\circ}}$.
In contrast, the PnP algorithm and CA-RSSR almost cannot be implemented
for $\Psi_{c}=10{^{\circ}}$. In addition, the CR of the three algorithms
decrease slightly with large FoV since the power of shot noise increases
\cite{cui2010line}.

\vspace{-0cm}

\subsection{\label{subsec:Accuracy-Performance}Accuracy Performance Of R-P3P}

In this subsection, we evaluate the accuracy performance of R-P3P
under the influence of LEDs orientation, the image noise and the
distance between the camera and the PD on the receiver.

1) Effect of the LED orientation

\begin{figure}[htbp]
\centering{}\includegraphics[scale=0.7]{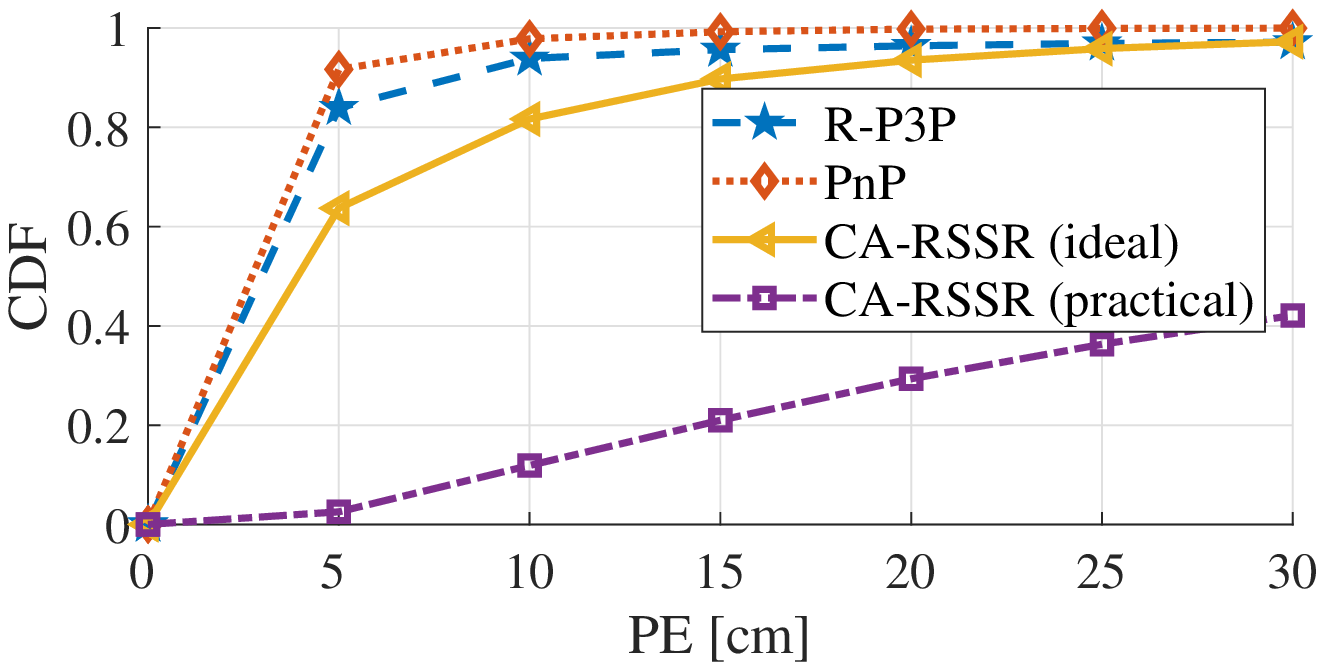}\caption{\label{fig:compare-of angle error}The comparison of 3D-positioning
accuracy performance among R-P3P, CA-RSSR and the PnP algorithm with
a random tilt angle $\theta$ of LEDs.}
\end{figure}

We first evaluate the effect of LEDs orientation on 3D-positioning
accuracy of R-P3P, CA-RSSR and the PnP algorithm. CA-RSSR requires
the LEDs to face vertically downwards, which may be challenging to
satisfy in practice. Therefore, two cases are considered for CA-RSSR:
the ideal case where the LEDs face vertically downwards,
and the practical case where the LEDs tilt with a random angle perturbation
$\theta\leq5{^{\circ}}$. In contrast, R-P3P and the PnP algorithms
can be implemented in the two cases, and thus only the practical case
is considered for them. The accuracy performance is represented by
the cumulative distribution function (CDF) of the PEs. As shown in
Fig. \ref{fig:compare-of angle error}, R-P3P achieves 80th percentile
accuracies of about $5\:\mathrm{cm}$, which is almost the same with
the PnP algorithm. This implies that the probability of the situations
that more than one groups of $\phi_{i}$ ($i\in\left\{ 1,2,\ldots,K\right\} $)
comply (\ref{eq:35}) or no group of $\phi_{i}$ ($i\in\left\{ 1,2,\ldots,K\right\} $)
complies (\ref{eq:35}) is very low. Therefore, although (\ref{eq:35})
is not strict in theory, the accuracy of R-P3P is close to that of
the PnP algorithm using less LEDs. Besides, CA-RSSR achieves 80th
percentile accuracies of about 10 cm for the ideal case. However,
the practical case of CA-RSSR presents a significant accuracy decline
compared with the ideal case of the CA-RSSR. Thus, a slight LEDs orientation
perturbation can impair the accuracy significantly for the CA-RSSR.

\begin{figure}
\centering\includegraphics[scale=0.7]{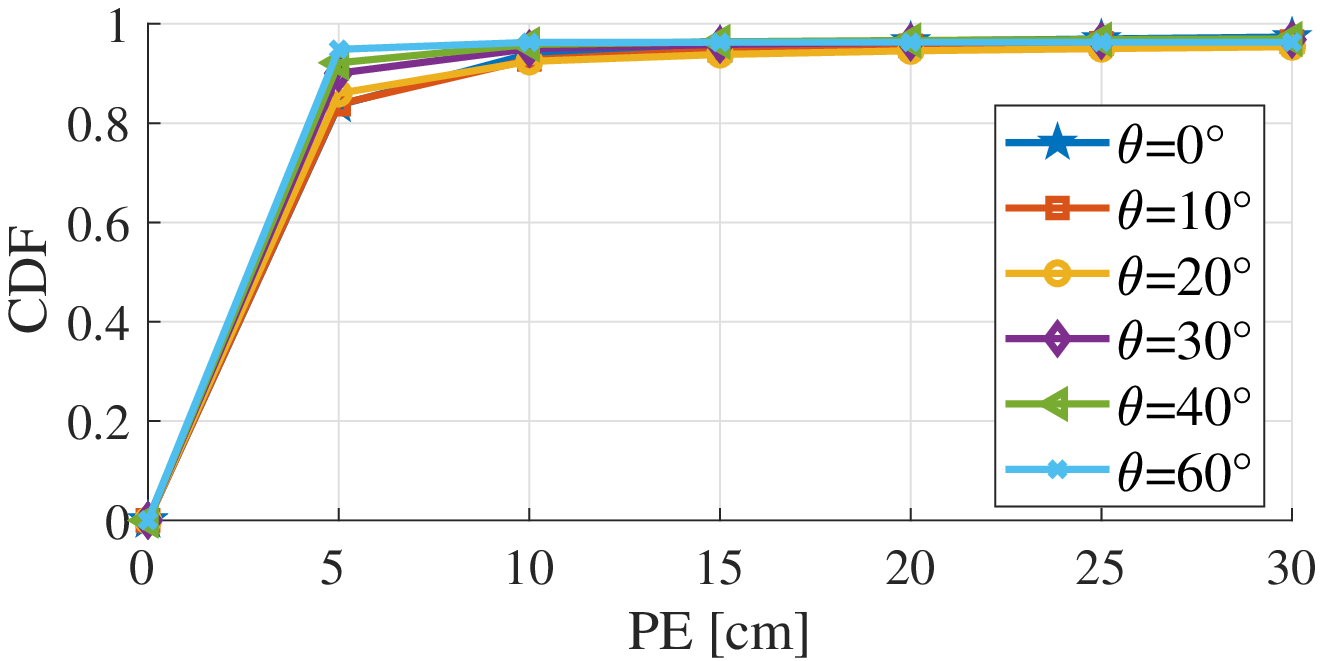}\centering
\caption{\label{fig:LED tilt angle}The effect of the tilt angle of LEDs on
the accuracy performance of R-P3P.}
\end{figure}

Then, we evaluate the 3D-positioning accuracy of R-P3P with varying tilt
angles of LEDs. The performance is represented by the CDF of PEs, given
$\theta=0{^{\circ}}$, $10{^{\circ}}$, $20{^{\circ}}$, $30{^{\circ}}$,
$40{^{\circ}}$ and $60{^{\circ}}$. As shown in Fig. \ref{fig:LED tilt angle},
R-P3P can achieve 80th percentile accuracies of less than 5 cm for
all $\theta$. Therefore, R-P3P can be utilized widely in the scenarios
where the LEDs are in any orientation. Besides, the accuracy of R-P3P
increases slightly as the tilt angle of LEDs increases since the irradiance
angles decrease which further improves the received signal power.

\begin{figure}[htbp]
\centering\includegraphics[scale=0.7]{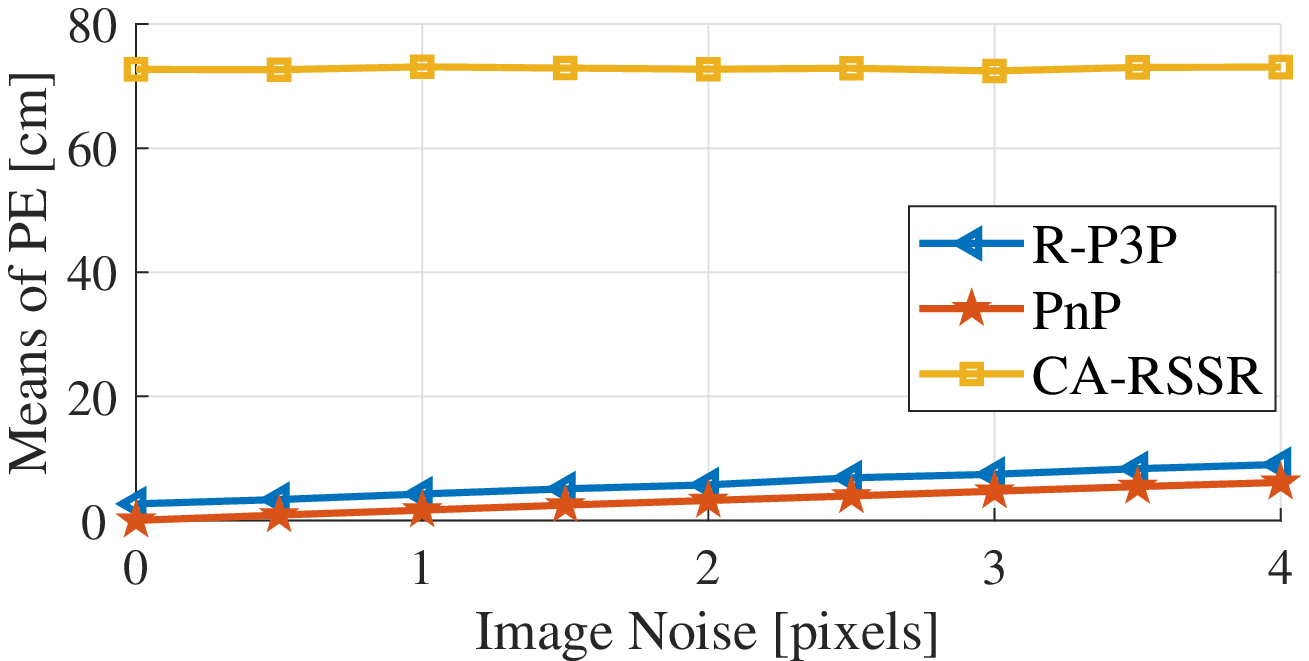}\centering \caption{\label{fig:The-influence-of image noise}The comparison of the effect
of the image noise on 3D-positioning accuracy performance among R-P3P,
CA-RSSR and the PnP algorithm under the case where LEDs tilt with
a random angle perturbation $\theta\protect\leq5{^{\circ}}$.}
\end{figure}

\vspace{-0cm}

2) Effect of the image noise

Since R-P3P also exploits visual information, we then evaluate the
effect of the image noise on the accuracy performance of R-P3P, CA-RSSR
and the PnP algorithms for 3D positioning under the case where the
LEDs tilt with a random angle perturbation $\theta\leq5{^{\circ}}$.
The image noise is modeled as a white Gaussian noise having an expectation
of zero and a standard deviation ranging from 0 to $4$ pixels \cite{zhou2019robust}.
The mean of PEs that are affected by the image noise are shown in
Fig. \ref{fig:The-influence-of image noise}. As shown in Fig. \ref{fig:The-influence-of image noise},
the accuracy performance of R-P3P closes to that of the PnP algorithm
and is much better than that of CA-RSSR. For R-P3P, the means of PEs
increase from 3 cm to 10 cm with the increasing of the image noise.
For the PnP algorithm, the means of PEs increase from 0 to 9 cm.
In contrast, for CA-RSSR, the means of PEs keeps at about 72 cm.
\begin{center}
\vspace{-0cm}
\par\end{center}

3) Effect of the distance between the PD and the camera

\begin{figure}
\centering \includegraphics[scale=0.65]{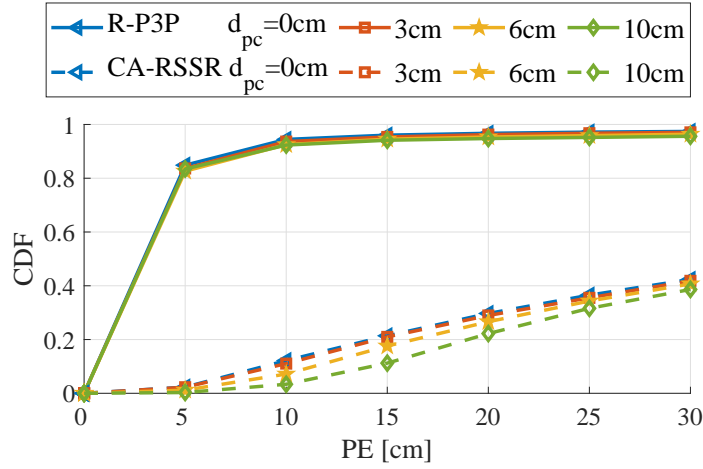} \centering \caption{\label{fig:dpc}The comparison of 3D-positioning accuracy performance
for CA-RSSR and R-P3P with varying distances between the PD and the
camera under the case where LEDs tilt with a random angle perturbation
$\theta\protect\leq5{^{\circ}}$.}
\end{figure}

Since R-P3P exploits the PD and the camera simultaneously, we then
evaluate the effect of the distance between the PD and the camera,
$d_{\textrm{PC}}$, on the accuracy performance of R-P3P. We compare
CA-RSSR and R-P3P on 3D-positioning performance with varying $d_{\textrm{PC}}$
under the case where the LEDs tilt with a random angle perturbation
$\theta\leq5{^{\circ}}$. This performance is represented by the CDF
of the PEs with $d_{\textrm{PC}}=0\;\mathrm{cm}$, $1\;\mathrm{cm}$, $3\;\mathrm{cm}$,
$6\;\mathrm{cm}$ and $10\;\mathrm{cm}$. In particular, $d_{\textrm{PC}}=0\;\mathrm{cm}$
indicates the ideal case that the PD and the camera overlap. As shown
in Fig. \ref{fig:dpc}, compared with CA-RSSR, R-P3P can achieve better
performance. In specific, R-P3P can achieve 80th percentile accuracies
of about 5 cm regardless of $d_{\textrm{PC}}$. In contrast, CA-RSSR
can only achieve 40th percentile accuracies of about 30 cm for all
$d_{\textrm{PC}}$. As we can observe from Fig. \ref{fig:dpc}, $d_{\textrm{PC}}$
has little effect on positioning accuracy of R-P3P.
This means that R-P3P can
be widely used on devices with various $d_{\textrm{PC}}$.

\subsection{Computational Cost}

\begin{figure}
\centering \includegraphics[scale=0.70]{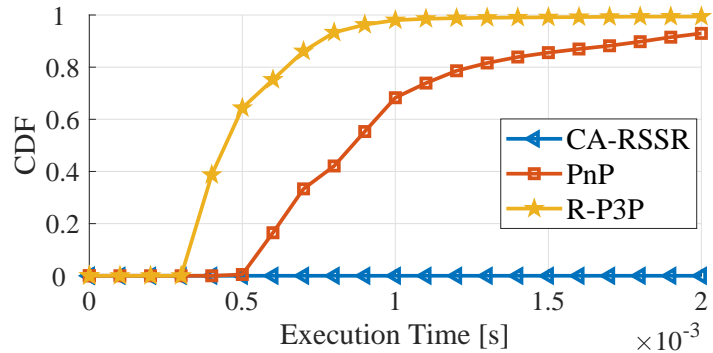}\caption{\label{fig:time cost}The computational cost of 3D-positioning for
R-P3P, CA-RSSR and the PnP algorithm.}
\end{figure}

In this subsection, we compare execution time of R-P3P, CA-RSSR and
the PnP algorithm for 3D positioning to evaluate the computational
cost performance \cite{kneip2011novel}\cite{lim2015ubiquitous}.
To have a fair comparison, all algorithms have been implemented in
Matlab on a $1.6\mathrm{GHz}\times4$ Core laptop. The experiment consists
of $10^{5}$ runs. The results are shown in Fig. \ref{fig:time cost}.
Since R-P3P estimates the position of the receiver by the LLS method,
the computational cost of R-P3P is the lower than that of CA-RSSR,
and the execution time of it is shorter than $0.001\:\mathrm{s}$ for
almost 100\% of the $10^{5}$ runs. Considering a typical indoor walking
speed 1.3 m/s, the execution delay of R-P3P only causes 0.2 cm positioning
error, which is acceptable for most applications. Besides, the computational
cost of the PnP algorithm is over $0.002\:\mathrm{s}$ for about 90\%
of the $10^{5}$ runs, which means the computational cost of R-P3P
is less than 50\% of that of the PnP algorithm.

\vspace{-0cm}

\section{CONCLUSION}

\label{sec:CONCLUSION}We proposed a novel indoor positioning algorithm
named R-P3P that simultaneously utilizes visual and strength information.
Based on the joint of visual and strength information, R-P3P can mitigate
the limitation on LEDs orientation. Besides, R-P3P can achieve better
accuracy performance than CA-RSSR with low complexity due to the use
of the LLS method. Furthermore, R-P3P requires less LEDs than the
PnP algorithm. Simulation results indicate that R-P3P can achieve positioning accuracy
within 10 cm  over 70\% indoor area  with low complexity regardless of
LEDs orientations. Therefore, R-P3P is a promising
indoor VLP approach, which can be widely used in the scenarios where
the LEDs are in any orientation. In the future, we will experimentally
implement R-P3P and evaluate it using a dedicated test bed, which
will be meaningful for future indoor positioning applications.

\vspace{-0cm}
 % -------------------------------------------------------------------------

%\bibliographystyle{plain}
%\bibliography{main_tex7}\bibliographystyle{plain}
 \bibliographystyle{IEEEbib}
\bibliography{BL_abbr_all}

%%%%%%%%%%%%%%%%%%%%%%% References %%%%%%%%%%%%%%%%%%%%%%%%%
%%%%%%%%%% If using BibTeX:
%\bibliography{sample}

%%%%%%%%%% If preparing manually:
% \newcommand{\enquote}[1]{``#1''}

% \bibitem{Zhang:14}
% Y.~Zhang, S.~Qiao, L.~Sun, Q.~W. Shi, W.~Huang, L.~Li, and Z.~Yang,
%   \enquote{Photoinduced active terahertz metamaterials with nanostructured
%   vanadium dioxide film deposited by sol-gel method,}
%   {\protect\JournalTitle{Optics Express}} \textbf{22}, 11070--11078 (2014).

% {Optical Society}, \enquote{{OSA Publishing},}
%   \url{http://www.osapublishing.org}.

% \bibitem{FORSTER2007}
% P.~Forster, V.~Ramaswamy, P.~Artaxo, T.~Bernsten, R.~Betts, D.~Fahey,
%   J.~Haywood, J.~Lean, D.~Lowe, G.~Myhre, J.~Nganga, R.~Prinn, G.~Raga,
%   M.~Schulz, and R.~V. Dorland, \enquote{Changes in atmospheric consituents and
%   in radiative forcing,} in \enquote{Climate Change 2007: The Physical Science
%   Basis. Contribution of Working Group 1 to the Fourth assesment report of
%   Intergovernmental Panel on Climate Change,}  S.~Solomon, D.~Qin, M.~Manning,
%   Z.~Chen, M.~Marquis, K.~B. Averyt, M.~Tignor, and H.~L. Miler, eds.
%   (Cambridge University Press, 2007).

% \end{thebibliography}

\end{document}